\documentclass[12pt]{article}
\usepackage[dvips]{graphicx}
\begin{document}

\begin{center}   {A Dark Matter Signature for Condensed Neutrinos}
\end{center}
\vspace{.2in}
\begin{center}
P.D. Morley \\  
System of Systems Analytics, Inc. \\ 11250 Waples Mill Road \\ Suite 300 \\Fairfax, VA 22030-7400 \\
{\it E-mail address:} peter3@uchicago.edu \\
\vspace{0.25in}
and \\
\vspace{0.25in}
 D. J. Buettner \\
Aerospace Corporation \\P. O. Box 92957 \\Los Angeles, CA 90009-2957 \\
{\it E-mail address:}  Douglas.J.Buettner@aero.org
\end{center}

\begin{abstract}
We derive the signature for condensed neutrino objects (CNOs) as the primary source of Dark Matter. Restricting our source data to minimize systematic errors, we find that by just using weak lensing data and Sunyaev-Zel'dovich data, that there may be a weak CNO signature.
\end{abstract}

\newpage

\tableofcontents

\section{Introduction}
Dark Matter has a long history\cite{Web}. However, we can reduce the discussion to the main attributes a theory must satisfy in order to explain astrophysical data. We maintain that the equation of state for whatever particle or particles that makes up Dark Matter is the key concept.

The first criteria is that Dark Matter must be able to support itself from gravitational collapse. The second criteria is that a Dark Matter object must be big enough to encompass a whole cluster of ordinary galaxies. These two criteria severely limit what particles Dark Matter could be made up of. If the Dark Matter particle is a boson, it must support itself through thermal pressure such as the ideal equation of state. If this were not so, it would condense into the lowest ground state and become a black hole. The characteristic scale of objects supporting themselves from thermal pressure are stars, but they also need an intrinsic energy source. No star can survive its own gravitational pull using thermal pressure alone unless it has intrinsic internal energy generation. It would seem then that bosons as Dark Matter have significant fundamental physics issues. 

The remaining Dark Matter particles will be fermions. Now, there are two ways such a Dark Matter object can preserve itself from its own gravity field: the first method is the same as a boson (the thermal equation of state), and the second method is from degeneracy pressure. The first mechanism suffers from the same problem as the last discussion because using thermal pressure for stability against gravitational collapse does not depend on the spin of the particle. The second method of degeneracy pressure, occurring in neutron stars (nucleons) and white dwarfs (electrons) is critically dependent on the mass of the fermion. Degenerate objects made up of fermions become increasingly smaller as the mass of the fermion increases. Thus, in order to have a Dark Matter object large enough to embed an ordinary cluster of galaxies within it, the fermion particle mass has to be much smaller than the mass of the electron. 

The third issue here is data integrity: when reducing raw data to processed data, assumptions nearly always have to be made, such as hydrostatic equilibrium for x-ray cluster gas, or neutrino degrees of freedom contribution to the cosmological black body pressure (meaning that the neutrinos really are in thermodynamic equilibrium when they uncouple). These assumptions, if incorrect, corrupt the processed data.

\section{Two Classes of Corrupted Data in Science}
Data corruption is a very real problem in science. Nature magazine has recently published evidence\cite{Nat} that the majority of biomedical experimental data is not reproducible. We call this `Corrupted Data', or CODA in this paper\footnote{We exclude deliberate falsification of data and discuss only true data errors.}. CODA also exists in the physics and astronomy community. CODA can be further classified as class I: the raw data itself is corrupted, or class II: the raw data is fine, but the processed final data is corrupted. A notable example in physics of a class I CODA is the CERN OPERA experiment that recorded neutrinos moving faster than light in a vacuum\cite{OPERA}. This was a `10 sigma discovery', meaning that the purely statistical chance for all their recorded neutrinos moving faster than light in a vacuum was in the extremely rare tenth sigma probability occurrence. However, OPERA had a systematic error that made the statistical ten sigma error irrelevant. 

Class II CODA can be extremely difficult to identify because the reasons for it can be wrong assumptions in processing the raw data, which the scientists believed were true\footnote{The analogy in mathematics of such a class II CODA would be the following: a theorem proof where one of the steps is wrongly assumed true.}. In astronomy, CODA is famously linked with cosmology\cite{Hoyle}. From nuclear and particle physics, we list three CODA situations from a small four-year window in the middle '70s.

\begin{itemize}
\item {\it Large Z Muonic Atoms} \\
Reference\cite{Dix} created large Z muonic atoms and measured the cascade x-ray energies that occur as the muon descended the excited energy staircase. By concentrating on large orbital angular momentum states, they avoided the problem of knowing the protonic charge distribution in large Z atoms. They were looking for violations of quantum electrodynamics (QED) because the two-loop vacuum polarization term is important\cite{Jen}. Having found a discrepancy between QED and their data, they declared the former broke down. This experiment was repeated by a different group\cite{Tau} who got agreement with the QED theoretical calculations and who concluded that in reference\cite{Dix} `Systematic errors were seriously under-estimated'.

\item {\it High-y Anomaly in Anti-neutrino Scattering} \\
In 1974 reference\cite{Aub} reported evidence of a `high-y anomaly' in anti-neutrino scattering which called into question weak interaction charge symmetry, left-handed currents and Bjorken scaling. What makes this CODA so interesting is that it spawned `copy-cat' results\cite{Ben1, Ben2, Bar}. Anti-neutrino scattering was done by a later group\cite{Holder} who got agreement with the electro-weak theory and the parton model and who concluded `There is no high-y anomaly'.

\item{Violation of Unitarity in K$^{0}_{L} \rightarrow \mu^{+}+\mu^{-}$ } \\
Unitarity in quantum mechanics basically means conservation of probability; that is, summing all possibilities of branching ratios for a particle decay should equal 100\%. In rare instances, unitarity sets a {\it lower} bound to a branching ratio as it does\cite{Sig} for the decay K$^{0}_{L} \rightarrow \mu^{+}+\mu^{-}$. An experiment testing unitarity was done by reference\cite{Clark} who found too few di-muon decays of K$^{0}_{L}$ and conservation of probability was in doubt. An experiment was undetaken\cite{Car} to check the claim of\cite{Clark}. The results of\cite{Car} were in good agreement with the unitarity bound and in sharp disagreement with the claim, deciding that the probability that unitarity was violated is `less than 0.06\%'\footnote{As recently as 2016, a similar situation is occurring for neutrino mixing where unitarity is again called into question\cite{CH}.} .
\end{itemize}

In these examples, systematic errors caused both class I and class II CODA. In this paper, we critically examine three data sets in astronomy: Planck satellite data analysis, N-body gravitational collapse simulations and galaxy cluster x-ray data.

\section{Planck Satellite Data Analysis Assumption}
An assumption for the Planck data analysis\cite{Pla} is that the cosmological neutrinos are in full thermal equilibrium prior to neutrino decoupling and posits equations for the radiation field thermodynamic temperature $T_{\gamma}$ and density $\rho_{\gamma}$ as (the subscript $\nu$ is used to indicate neutrino quantities):
\begin{eqnarray}
T_{\gamma} &  \stackrel{{\rm ?}}{=} & (\frac{11}{4})^{\frac{1}{3}} T_{\nu} \\
\rho_{\nu}  & \stackrel{{\rm ?}}{=}  &  N_{{\rm eff}}\frac{7}{8}(\frac{4}{11})^{\frac{4}{3}}\rho_{\gamma}
\end{eqnarray}
Are these assumed equations correct? Current experimental evidence\cite{Ol} suggests that neutrinos have masses $<$ 2 eV/c$^{2}$, and small neutrino mass mixings (implying near neutrino mass degeneracy). No evidence of the Majorana neutrino type has been found to date. Thus, if the present three neutrino flavors are indeed Dirac neutrinos, they will possess a permanent magnetic dipole moment (see for instance reference\cite{Abo}). Whether such a dipole moment is anomalous or not, the early universe will induce non-equilibrium neutrino power radiation losses\cite{MB1}. Relativistic magnetic moments radiate as $\sim \gamma^{8}$, compared to the charged particle Li\'{e}nard result $\sim \gamma^{6}$, where $\gamma$ is the usual relativistic factor $1/\sqrt{1 -\frac{v^{2}}{c^{2}}}$. It is important to note that a neutrino's radiation loss is not so much dependent on the scale of Big Bang magnetic fields as on their fluctuations; that is, the continuous flipping back and forth of the dipole is a major determiner of their spontaneous radiation emission power loss. Because the initial Big Bang will produce high magnitude, highly turbulent (fluctuating) plasmas\cite{Gra,Wag}, there will, in principle, be no limit to a neutrino's power loss.

Keeping neutrino degrees of freedom in the cosmological black-body microwave data reduction leads to tension between the Planck consortium cosmological parameter values and their estimates from other independent methods\cite{Zh4}. If neutrinos are removed from the cosmological microwave data, some other degree of freedom must replace their contributions to the perturbations on the acoustic peaks\footnote{That is, their effect on the multipole moments.}. Reference\cite{Ad} shows that a cosmic magnetic field has the same contribution to the cosmic microwave background as kinetic degrees of freedom. By replacing neutrino degrees of freedom with magnetic field degrees of freedom, it is expected that the cosmological microwave data will then lead to a value of the Hubble constant that will be in line with values derived from its direct determination. As an added benefit, the early universe magnetic fields will lead to remnant, vestigial intergalactic magnetic fields seen today\cite{Es}. The assumption that cosmological neutrinos in the early universe are in thermodynamic equilibrium with baryonic matter cannot be maintained if neutrinos are only of the Dirac type.

If neutrinos lose their energy through radiation power losses in the early universe, cold neutrinos will condense\footnote{The history of neutrino condensation is a long one beginning in 1973 with the suggestion of reference\cite{CM}. In reference\cite{MB2}, an account is given of the 1980s revival with heavy neutrino masses.}. CNOs satisfy all of the requirements for Dark Matter by virtue of the modern theory of the electro-weak force and the present neutrino mass bound: their hydrostatic equilibrium produces sizes that are millions of parsecs in radius and thousands of times more massive than the Milky Way, in a single object. They are so large they easily embed a galaxy or a cluster of galaxies. If that were to happen, the embedded galaxies would undergo simple harmonic motion\footnote{Superimposed on their peculiar motions.} within the CNO\cite{MB2}. A CNO would be totally transparent to light: one can see right through them. The non-equilibrium neutrino radiation loss and the initial phases of condensation should appear as hot spots in the cosmological microwave radiation.  

A significant difference between charged matter and neutrino matter is the following: the Big Bang will have created equal parts of matter and anti-matter, but the near-annihilation of charged matter by its anti-matter counter-part will produce a reduced baryonic matter density (`baryon asymmetry problem'). However, since neutrinos hardly annihilate\footnote{Specific cross sections are discussed in reference\cite{MB2}.} with their anti-neutrinos, the present day universe should have essentially the original numbers of neutrinos and anti-neutrinos. This difference of interaction causes a large energy density disparity between charged matter and neutrino matter. Thus CNOs satisfy the Dark Matter energy density prominence over charged matter.

\section{The Dark Matter Signature for Condensed Neutrinos}
The equation of state for a CNO reference\cite{MB2} is exceedingly concise: a relativistic Fermi gas with degeneracy\footnote{Neutrino mixing means there are no well defined leptonic flavors - equal  numbers of anti-neutrinos and neutrinos means having net leptonic charge zero.} of 6. Thus, a hydrostatic solution for condensed neutrinos has just two parameters: $m_{\nu}$ (hereafter just $m$) the common neutrino and anti-neutrino degenerate mass (called the `common degenerate neutrino mass') and the Fermi momentum $p_{F}$ at the center of the CNO, $p_{F}(0)$. These quantities enter in the ratio $x \equiv p_{F}/$mc where c is the speed of light. The relevant astronomical CNO configurations have $x \ll 1$.

\subsection{CNO Solutions}
In Fig. 1, we present CNO solutions for $m = {\rm 1.0\; eV}/c^{2}$ mass. This is the signature of a CNO: higher mass CNOs have smaller radii. In fact, all Fermi degenerate objects (neutron stars and white dwarfs) have the property that the more massive the object is, the smaller in size it becomes\footnote {If this were not so and the graph had {\it positive} slope, then a perturbation inward would support less mass and the degenerate object would be unstable to radial perturbations.}. This stands in stark contrast to hot gases, plasmas and iron-silicate equations of state, all of which have positive gradient mass versus radius. It presents a unique identifier for degenerate matter.

\begin{figure}
\includegraphics[scale=0.8]{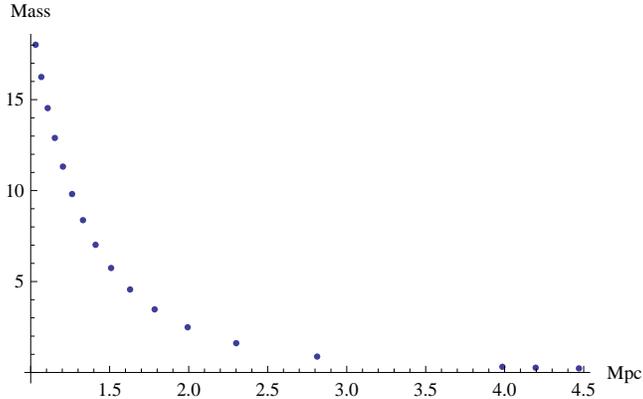}
\caption{Numerical soutions (CNO signature) for  $mc^{2}$ = 1.0 eV, where the radius is in Mpc and the mass is in $1\times10^{14}$ solar masses. Both the CNO radius and mass scale as 1/$m^{2}$ where $m$ is in eV/c$^{2}$ units. }
\label{Fig. 1}
\end{figure}

This signature has the approximate parameterization (see Appendix) for general  $m$ 
\begin{equation}
M(R) \simeq \frac{1.97462    \times 10^{15} M_{\odot}    }{R^{3}m^{8}}
\end{equation}
where $M(R)$ is the mass of the CNO having radius $R$ in units (Mpc) and $m$ is the common degenerate neutrino mass in units eV/c$^{2}$. Eq(3) is not exact, but reasonable.

\subsection{Degenerate Matter Density Profiles}
We generate the density profiles for astronomically relevant solutions in Fig. 2.
\begin{figure}
\includegraphics[scale=0.75]{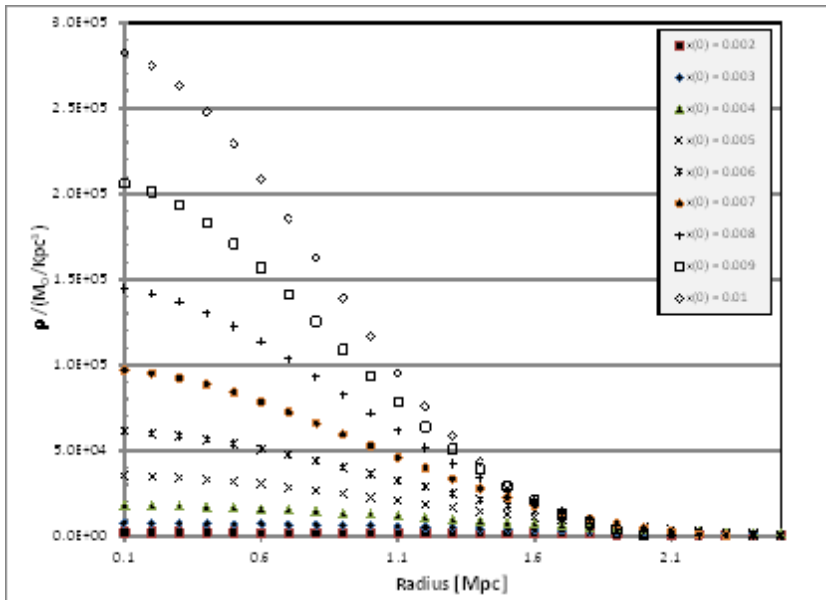}
\caption{Astronomically relevant degenerate neutrino assemblies with the common $\nu$ mass 0.8 eV/c$^{2}$.}
\label{Fig. 2}
\end{figure}

\section{N-body Gravitational Collapse Simulations}
Physics based numerical N-body simulations have been used in attempts to identify the gravitational contribution to dwarf galaxy 'halos' and clusters of galaxies 'halos' from Cold Dark Matter\cite{Nav}. However, if Dark Matter is degenerate, these numerical profiles would then be based on an incorrect assumption. To see this, the NFW profile is derived as
\begin{equation}
\rho(r) = \frac{\delta_{c} \rho_{c}}{(r/r_{s})(1+r/r_{s})^{2}}
\end{equation}
where $\delta_{c}$, $\rho_{c}$ and $r_{s}$ are parameters. The highly peaked nature of this density function is incorrect for degenerate Fermi matter, because degenerate Fermi matter has a finite Fermi momentum at the center (Fig. 2). In actual fact, N-body simulations of classical matter will never give degenerate matter unless it is included as a required constraint. Monte Carlo simulations of degenerate matter is possible if this constraint is implemented\cite{RLE}. 

\section{The X-ray Temperature Method}
Most of the available Cold Dark Matter virial mass and radii data for galaxies are derived from the temperatures of X-rays obtained from X-ray satellites.  In order for this to happen, hydrostatic thermodynamic equilibrium is assumed. There are potential statistical errors associated with deriving masses and radii of Dark Matter in this manner, which we discuss in this section. However, the method turns out to have a bias which is only evident when one plots the totality of the data.
\begin{itemize}
\item Dark Matter does not radiate so the center of the Dark Matter object is difficult to determine. The embedded galaxies may be in simple harmonic motion\cite{MB2} and thus the center of the galaxy cluster need not correspond to the center of the Dark Matter's distribution. Displaced Dark Matter centers will then show an erroneous, spurious `cooling' center (smaller keV temperature), leading to an error in the actual radius of the Dark Matter object.
\item The assumption that the hot x-ray cluster gas is in thermal equilibrium leads to the suspect equation\cite{Zh9}:
\begin{equation}
\frac{1}{\mu m_{p} n_{e}(r)}\frac{d[n_{e}(r)T(r)]}{dr} \stackrel{{\rm ?}}{=} -\frac{GM(<r)}{r^{2}}
\end{equation}
where $m_{p}$ is the proton mass, $n_{e}$ is the electron density and $\mu$ is the average molecular weight for hydrogen. The hot gas has a long cooling time\cite{MT} and is also executing simple harmonic motion in the Dark Matter potential well. There is no requirement at all for stability against gravitational collapse. Reference\cite{Ma} finds that the x-ray emitting gas in galaxy clusters is not in hydro-static equilibrium.
\item The gas is hot because it fell down a potential well so the following equation may look attractive:
\begin{equation}
\frac{3}{2}kT_{{\rm MAX}} \stackrel{{\rm ?}}{=}  \frac{Gm_{p}M_{{\rm virial}}}{R_{{\rm virial}}} + \; {\rm corrections}
\end{equation}
\end{itemize}
However, the gas may have had additional heat injection\cite{MT} (quasars, supernovae, active galactic nuclei, etc.) and/or had significant kinetic energy before it fell into the gravitational potential well; the latter could be epoch-dependent. While authors attempt to correct for these effects, these unknown correction terms likely lead to erroneous results for Dark Matter virial mass.

These potential errors may be completely over-shadowed by the hydrostatic thermodynamic equilibrium assumption. In Fig. 3 we plot the data from virial masses and radii from the Fermi\cite{S} and Chandra satellites\cite{Sc}.

\begin{figure*}
\includegraphics[scale=0.6]{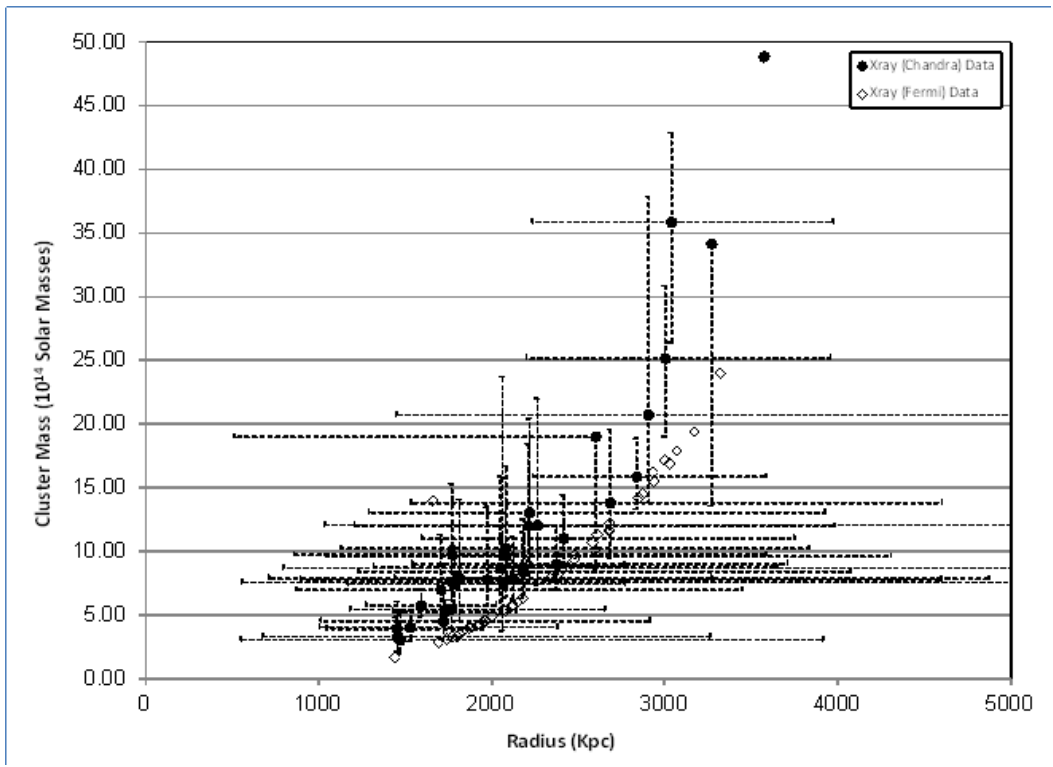}
\caption{Virial mass and radii of galaxy clusters using the Fermi and Chandra satellites. }
\label{Fig. 3}
\end{figure*}

The error bars for the Fermi satellite data are not shown for ease of viewing\footnote{The error bars are statistical errors, but there may be a systematic error as well which we discuss above.}. Also, the paper of reference\cite{Sc} has been corrected due to an important typographical mistake\footnote{Private communication from R. Schmidt.}. Figure 3 does not show the trend of large virial mass, small radius that is the smoking gun of Fermi degenerate matter; instead, it shows large virial mass with large radius. {\it This is the equation of state for a plasma.} The input assumption of a hot gas in hydrostatic thermodynamic equilibrium producing an output result of a plasma equation of state is a spectacular example of circular reasoning. Circular reasoning is the systematic error of the processed data. What should have been the assumption for the hot gas? Reference\cite{MB2} showed that galaxies embedded in Dark Matter actually undergo simple harmonic motion (SHM). Thus, the hot gas within the same gravitational potential well may also be expected to undego the same motions. 

For these reasons, we turn to the more direct methods of obtaining (Cold) Dark Matter virial mass and radii. Our literature search left us with only a few weak lensing candidates and a single Sunyeav-Zel'dovich method candidate. We elected to not use strong lensing data in order to minimize the potential source of error from attempting to determine these virial values from an observed arc versus the entire ellipse.

\section{Comparison to Experimental Data}
The size of a CNO means that only clusters of galaxies will realistically be available as profilers. We are interested in the virial mass and the virial radius of galaxy clusters, and to see if the relation between them follows the CNO signature of degenerate Fermi matter, as shown in Fig. 1.  In view of the previous section, the safest data to use is weak lensing data which has weak coupling to assumed density profiles and the 
 Sunyaev-Zel'dovich method. Table 1 is the limited data set. The Hubble constant is parameterized as $H_{0}$ = 100h km/s/Mpc, with h ranging from 0.70 - 0.75. The variation of h is inconsequential compared to the error bars on the virial mass and radius. In Fig. 4, a fit is obtained using the Mathematica (R)\cite{Wo} NonLinearModelFit optimizing routine. For this sparse data set, $m$ is between $\sim 0.75-0.85$ eV/c$^{2}$ with a mean of $\sim 0.81$ eV/c$^{2}$. This is in direct conflict with the Planck satellite data analysis\cite{Pla} if neutrino degrees of freedom are allowed to contribute to the multipole moments of the cosmological microwave background.

\begin{table*}
    \begin{tabular}{cccc}
\multicolumn{4}{c}{} \\ \hline
cluster & mass ($10^{14}$ M$_{\odot})$ & radius (Mpc) & reference \\ \hline
\mbox{} & \mbox{} & \mbox{} & \mbox{} \\
MS2137-23 & 7.72$^{+.47}_{-.42}$ & 1.89$^{+.04}_{-.04}$ & \cite{Ga5} \\
Coma & 5.1$^{+4.3}_{-2.1}$ & 1.8$^{+.6}_{-.3}$ & \cite{Ga9} \\
A914 & 14.6$^{+8}_{-8}$ & 1.64$^{+.17}_{-.16}$ & \cite{IR}  \\
A1351 & 44$^{+18.6}_{-18.6}$ & 2.24$^{+.24}_{-.17}$ & \cite{IR} \\
A1576 & 53$^{+18.6}_{-18.6}$ & 2$^{+.13}_{-.28}$ & \cite{IR} \\
A1722 & 16$^{+8}_{-8}$ & 2.4$^{+.3}_{-.25}$ & \cite{IR}  \\
A1995 & 14$^{+5.3}_{-5.3}$ & 1.5$^{+.2}_{-.26}$ & \cite{IR} \\
A2390 & 13.3$^{+1.3}_{-10}$ & 2.53$^{+.13}_{-.13}$ & \cite{Dia} \\
MS1358 & 9.3$^{+1.3}_{-3}$ & 2.66$^{+.13}_{-.13}$ & \cite{Dia} \\
CL0024 & 5.3$^{+1.3}_{-5}$ & 2.53$^{+.13}_{-.13}$ & \cite{Dia} \\
A1689 & 16.9$^{+.3}_{-.28}$ & 2.6$^{+.15}_{-.15} $ & \cite{Br} \\
ACT-CL J0102-4915 `El Gordo' & 18.6$^{+5.4}_{-4.9}$ &  2.1$^{+.189}_{-.189}$ & \cite{Men}
\end{tabular}
\caption{Weak lens and Sunyaev-Zel'dovich data (`El Gordo') processed data.}
\end{table*}

\begin{figure}
\includegraphics[scale=0.63]{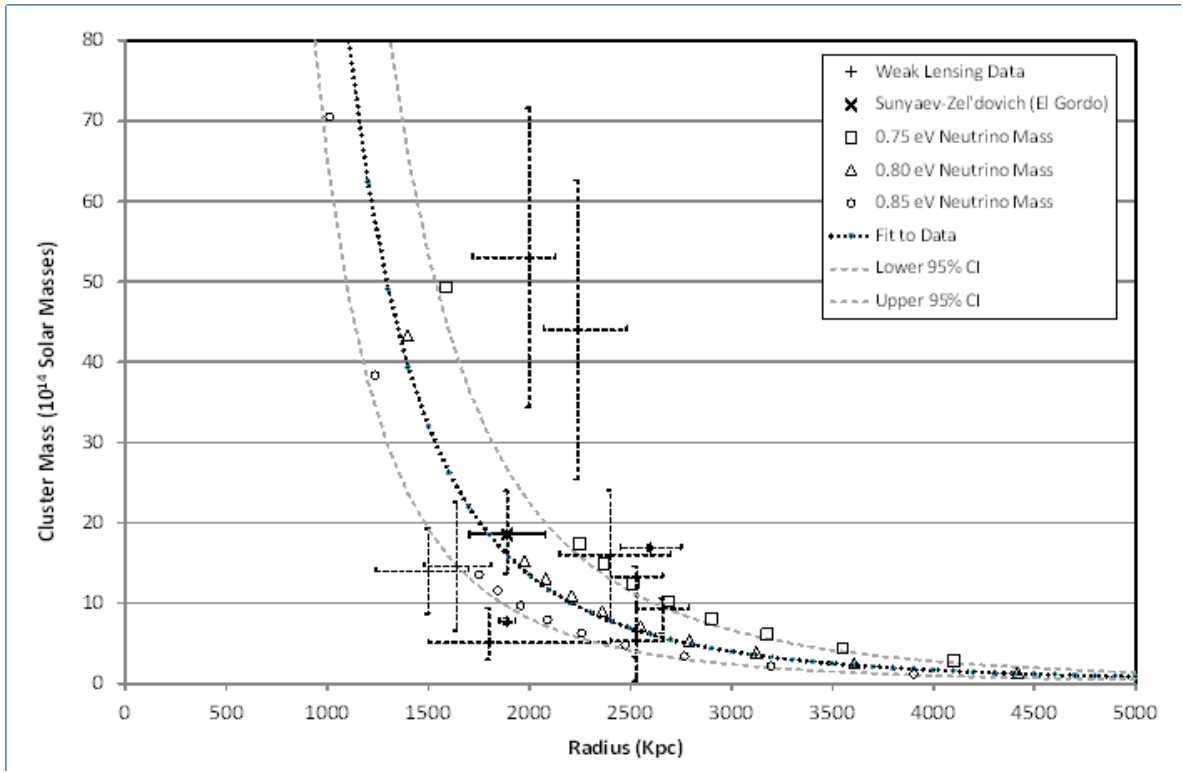}
\caption{Comparison of Table 1 virial mass and radius with super-imposed Eq(3) CNO ($m$) solutions.}
\label{Fig. 4}
\end{figure}

\section{Estimation of the Number of CNOs in the Universe}
Not surprisingly, CNOs repel each other due to the Pauli Exclusion Principle\cite{MB2}. Thus CNOs would exist as distinct separate entities\footnote{Possible bi-spherical structures may be possible.} and the possible number of CNOs in the universe is an interesting number. We can make a first order estimate here by assuming that the CNOs are the Dark Matter constituents and by neglecting possible interesting bi-spherical solutions\cite{MB2} which could lead to filament geometries. The present size\cite{wiki} of the universe ($4 \times 10^{86}$ cm$^{3}$) and its Dark Matter contribution\cite{wiki} (0.268) to the critical density ($1\times10^{29}$ gm/cm$^{3}$) then give the total mass for CNOs in gm: $1.07 \times 10^{57}$. Taking a typical CNO mass as $\sim 7 \times 10^{14}$ solar masses, the first order estimate of the number of CNOs in the universe $N_{CNO}$ is 
\begin{equation}
N_{CNO} \simeq 8 \times 10^{8} \sim \; {\rm 1 \; billion}
\end{equation}

\section{Gravitational Coupling of a CNO with embedded galaxies}
It was shown in reference\cite{MB2} that galaxies embedded within a CNO execute simple harmonic motion with periods of hundreds of millions of years. If a CNO contains galaxies in SHM, then there will be 
a gravitational back-reaction on the CNO. The resulting coupling will cause the CNO to go into its normal modes. The first normal mode will be a quadrupole breathing mode and in this section we calculate its period. To put the final answer in context, recall that great quakes within the earth cause this planet to ring like a bell: earth's first excited motion is also the quadrupole breathing mode, having a 55.55 minute period\cite{Mont} ($3\times10^{-4}$ Hz). This earth breathing mode and the CNO breathing mode are identical in concept.

The gravitational potential energy of the CNO is
\begin{equation}
\Omega({\cal R}) = -G\int_{0}^{{\cal R}} \frac{M(r)}{r}4\pi r^{2} \rho(r) dr
\end{equation}
where $\cal R$ is the CNO's radius, $M(r)$ is the mass within $r$ and $\rho(r)$ is the mass density. The neutrino and anti-neutrino particles experience a potential well made up of their own mutual gravitational attraction, so we can identify an effective potential energy as a function of radial coordinate $\xi$ with $\xi \leq {\cal R}$.
\begin{equation}
V(\xi) = - \Omega(\xi) = \frac{1}{2}M\omega^{2}\xi^{2}
\end{equation}
where $M$ is the mass of the CNO. The quadrupole breathing mode has radial frequency\cite{Recat} $2 \omega$. We are interested in astronomically relevant CNO configurations, so we evaluate the quadrupole breathing mode for the configuration\footnote{This is a typical CNO fitted galaxy cluster configuration having mass $M = 1.532\times10^{15}$ M$_{\odot}$  with radius 1.97 Mpc. Such attributes warrant CNOs the moniker `giants of the universe'.} $x(0) = 0.01$ with $m$ = 0.80 eV/c$^{2}$. In Fig. 5, we graph the potential function and make a parabolic fit\footnote{The potential function is not purely harmonic, so it has a complicated vibration spectrum.}.  Using Eq(9) and remembering that the quadrupole mode is $2\omega$, we have 
\begin{eqnarray}
2 \nu & = & 1.366\times10^{-17} \; {\rm Hz} \\
\tau({\rm quadrupole}) & = & 2.32\times 10^{9} \; {\rm years}
\end{eqnarray}
This is so large a period that if the Local Group were embedded within this CNO, it would have barely completed two oscillations since the birth of planet Earth\footnote{Recognizing that the CNO occupy the lowest possible frequencies in nature, while Earth's vibration lies in the middle part of the scale, then the high frequencies in nature are bracketed by the breathing mode of atomic nuclei of which the $^{164}Er$ value of $2\times10^{19}$ Hz is an example\cite{Ga}.}. The result of this calculation is that the CNOs are hardly perturbed by the embedded galaxies and gas inside it. However, the distortion one CNO causes on another CNO in close proximity is probably substantial and should be calculated in the future.

\begin{figure}
\includegraphics[scale=0.6]{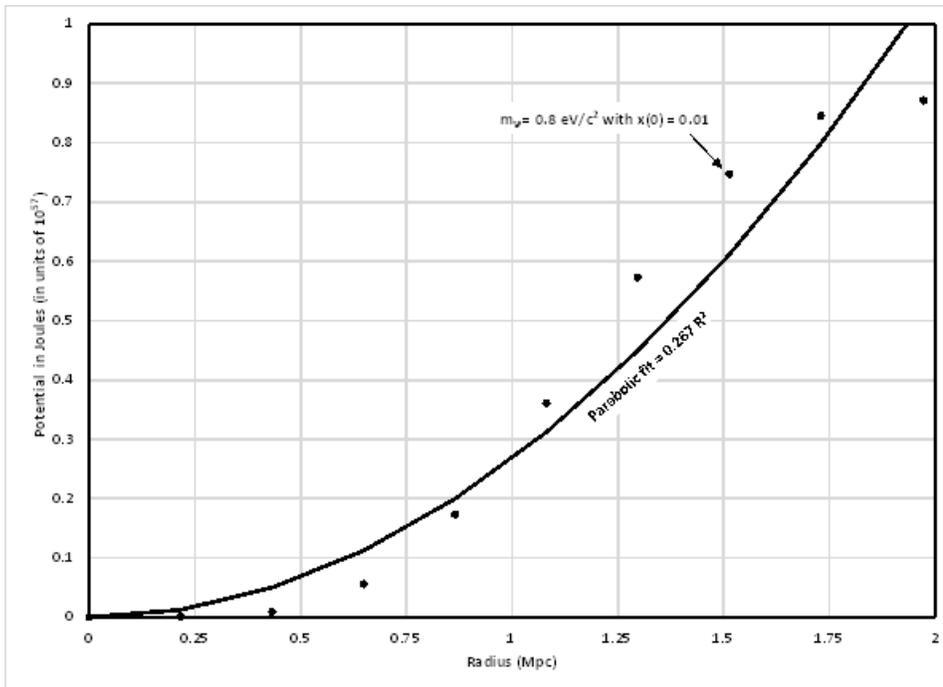}
\caption{Parabolic fit to the CNO potential function.}
\label{Fig. 5}
\end{figure}

\section{Conclusions}
Astronomers have known for years that there is an unidentified force affecting the rotational velocity behavior of galaxies. Reference\cite{MB2} showed that galaxies having spin axis not radially aligned with the center of a CNO have the additional observational effect of asymmetric rotational profiles\footnote{This means that if a galactic disk were folded in half onto itself, the opposite sides' rotational speeds are very different.}.

This paper considers the observational issues that arise if degenerate matter from condensed neutrinos is the primary source of (Cold) Dark Matter. As shown in reference\cite{MB1}, magnetic fluctuations from turbulent Big Bang magnetic fields may keep the neutrinos from achieving thermodynamic equilibrium and would lead to rapid neutrino cooling from magnetic spin-flipping. If they did in fact lose their energy and condense into degenerate objects, they would have a characteristic signature of increasing mass with smaller radii. These signatures were provided. As what should be an easy data fit exercise to determine if the theoretical distributions fit the experimental data, we find complications: the common degenerate neutrino mass we find is in conflict with published values from the Planck satellite consortium. We suggest in this paper that the early-universe neutrino degrees of freedom contribution to the acoustic multi-pole cosmic microwave background effects used by the Planck consortium be replaced by (the relativistic) early-universe magnetic degrees of freedom. We suggest that virial mass and radii derived from x-ray analysis of galaxy cluster gas suffers from bias. Finally, derived NFW Dark Matter profiles are inconsistent with degenerate Fermi matter. The N-body simulation community should consider performing simulations with degenerate matter density profiles as a constraint.

We have shown that the current estimated virial mass and radii of Dark Matter objects from weak lensing clusters of galaxies (with the one Sunyaev-Zel'dovich data point) has a positive signature for CNO. The signal is delicate and easily masked by assumptions about the observable hot plasma residing within the Dark Matter gravitational potential. Fitting a small data set gives the common degenerate neutrino mean mass as $\sim 0.81$ eV/c$^{2}$ bracketed by $\sim 0.75$ eV/c$^{2}$ and $\sim 0.85$ eV/c$^{2}$ at the 95\% confidence level. 

\newpage
\section{Appendix}
\appendix
\section{Derivation of Eq(3)}
The total mass $M_{T}(\zeta,m)$ and surface radius $R_{S}(\zeta,m)$ of a CNO is a function of the boundary condition $x = p_{F}/mc$ at the origin\footnote{The other boundary condition is $\frac{dx}{dr}=0$ @ r = 0.}  $x(0) = \zeta$, and the common (approximately degenerate mass) of the neutrinos and anti-neutrinos, $m$. For example, with $x(0) = 0.007$, and $m$ = 0.85 eV/c$^{2}$, the numerical solution is $M_{T} = 7.9528 \times 10^{14}$ M$_{\odot}$ and $R_{S} = 2.0881$ Mpc. From solving the differential equation for hydrostatic equilibrium, the functional dependence is noted\footnote{In these equations, $m$ is in the unit of eV/c$^{2}$.}
\begin{eqnarray}
M_{T}(\zeta,m) &  = & \frac{M(\zeta) }{m^{2}}  \\
R_{S}(\zeta,m) & = & \frac{R(\zeta)}{m^{2}}
\end{eqnarray}
For the case of $m$ = 1 eV/c$^{2}$, Figure 1 is
\begin{equation}
M(\zeta) = \frac{1.97462 \times 10^{15}  \; M_{\odot}    }{R(\zeta)^{3}} 
\end{equation}
where $R$ is in Mpc. Using Eqs[12,13], Eq[14] becomes
\begin{equation}
m^{2}M_{T}(\zeta,m) =  \frac{1.97462 \times 10^{15}}{[R_{S}(\zeta,m)m^{2}]^{3}} \; M_{\odot} 
\end{equation}
or
\begin{equation}
M_{T}(\zeta,m) =  \frac{1.97462 \times 10^{15}}{m^{8}[R_{S}(\zeta,m)]^{3}} \; M_{\odot} 
\end{equation}
which is Eq[3]. We can check this equation by putting in $\zeta = 0.007$, $m$ = 0.85 eV/c$^{2}$, and $R_{S}(0.007,.85) = 2.0881$ Mpc. Eq[16] gives $M_{T}(0.007,0.85) = 7.959 \times 10^{14} \;  M_{\odot}$, compared to the exact answer of 7.9528$\times 10^{14} \; M_{\odot}$.

\end{document}